\documentclass[10pt]{revtex4}
\usepackage{amssymb}
\usepackage{latexsym}
\usepackage{epsfig}
\usepackage{amsmath}
\begin{document}
\title{Holographic dark energy with the sign-changeable interaction term}
\author{M. Abdollahi Zadeh$^{1}$, A. Sheykhi$^{1,2}$\footnote{asheykhi@shirazu.ac.ir},
H. Moradpour$^2$\footnote{h.moradpour@riaam.ac.ir}}
\address{$^1$ Physics Department and Biruni Observatory, College of
Sciences, Shiraz University, Shiraz 71454, Iran\\
$^2$ Research Institute for Astronomy and Astrophysics of Maragha
(RIAAM), P.O. Box 55134-441, Maragha, Iran}

\begin{abstract}
We use three IR cutoffs, including the future event horizon, the
Hubble and Granda-Oliveros (GO) cutoffs, to construct three
holographic models of dark energy. Additionally, we consider a
Friedmann-Robertson-Walker (FRW) universe filled by a dark matter
(DM) and a dark energy that interact with each other through a
mutual sign-changeable interaction. Thereinafter, we address the
evolution of the some cosmological parameters, such as the
equation of state and dimensionless density parameters of dark
energy as well as the deceleration parameter, during the cosmic
evolution from the matter dominated era until the late time
acceleration. We observe that a holographic dark energy (HDE)
model with Hubble cutoff interacting with DM may be in line with
the current universe. Our study shows that models with the future
event horizon as the IR cutoff or the GO cutoff are in good
agreement with the observational data. In fact, we find out that
these obtained models can predict the universe transition from a
deceleration phase to the acceleration one in a compatible way
with observations. The three obtained models may also allow the
equation of state parameter to cross the phantom line, a result
which depends on the values of the system's constants such as the
value of the interaction coupling constant.
\end{abstract}

\maketitle

\section{Introduction}
Observations data from type Ia supernovae (SNIa)
\cite{Riess,Riess1,Riess2,Riess3}, the Large Scale Structure (LSS)
\cite{COL2001,COL20011,COL20012,COL20013} and the cosmic microwave
background (CMB) anisotropies \cite{HAN2000,HAN20001,HAN20002}
obviously signal us an accelerating universe at the present time.
From theoretical point of view, this expanding phase of the
universe is supported by an unknown source called dark energy (DE)
\cite{Roos}. The simplest model, which consists of a fluid with
positive energy density and negative pressure, is called
cosmological constant and suffers from some problems such as the
fine-tuning and the coincidence problems \cite{Roos}. Although,
many different models, such as quintessence
\cite{Wetterich1988,Wetterich19881}, phantom (ghost) field
\cite{Cald2002,Cald20021}, k-essence
\cite{Chiba2000,Chiba20001,Chiba20002}, Chaplygin gas
\cite{Kam2001,Kam20011}, agegraphic DE
\cite{Cai2007,Cai20071,Cai20072,Cai20073,Cai20074,Cai20075,Cai20076,Cai20077,
Sheykhi2009,Sheykhi20091,Sheykhi20092,Sheykhi20093,Sheykhi20094}
and ghost DE models \cite{Ohta,CaiG,shemov, SheyEb,SEM,SM} have
been proposed to justify the current accelerating phase of the
universe expansion, DE is still a confusing topic in modern
cosmology \cite{Rev1,Rev2,Rev3} and its nature and origin is an
unknown problem.

Following the developments in quantum theory of gravity, the
motivation to solve the DE problem, including its nature and
behavior, has arisen a lot of attentions \cite{Witten}. An example
of such effort is the HDE models
\cite{Cohen1999,Li2004,Hsu2004,Pavon2005,Sheykhi2010,Shey2011,
Weipil}. The area-entropy relation is the backbone of this
hypothesis claiming that black hole relates the short distance
cutoff to the long distance cutoff, which leads to an upper bound
for the zero-point energy density \cite{Cohen1999}. Based on this
theory, we have $\rho_D=3c^2M_p^2L^{-2}$ for the energy density of
HDE, where $c^2$ and $L$ are, respectively, a constant and the
infrared (IR) cutoff, and $M_{p}$ is the reduced Planck mass
\cite{Li2004,Hsu2004}. One of the basic assumption in the HDE
models is the suitable IR cutoff which leads to an accelerated
universe. The simplest choice for the IR cutoff is the Hubble
radius which is defined as, $L=H^{-1}$. However, it was argued
that Hubble radius cannot lead to an accelerating universe
\cite{Li2004}, unless the interaction between DE  and DM is taken
into account \cite{Pavon2005,Sheykhi2010}. In 2008, Granda and
Oliveros (GO) proposed a combination of the Hubble parameter  and
its first derivative as the IR cutoff of HDE models, namely
$L=(\alpha H^2+\beta \dot{H})^{-1/2}$ where $\alpha$ and $\beta$
are constant \cite{Granda2008,Granda2009}. This cutoff which is
known as the GO cutoff avoids the causality problem and solves the
coincidence problem and extensively investigated in the
literatures \cite{Jamil1,GODGP}.

Moreover, observations indicate that the possibility of a mutual
interaction between the DM and DE, which may solve the coincidence
problem, is not zero \cite{wang2016}. The simplest form of this
mutual interaction can be written as $Q=3b^2H\rho$ in which $b^2$
is a constant and $\rho$ can be the energy density of DM, DE or
even their sum \cite{wang2016}. Since $b^2$ is constant and $H$
and $\rho$ are positive quantities, the sign of the interaction
term is not changed during the cosmic evolution. However, recent
investigations confirm that the sign of the interaction between DM
and DE is changed during the history of the universe and in
particular in the redshift $0\cdot45\leq z\leq 0\cdot9$
\cite{Cai2010}. Clearly, the sign-changeable interaction term
cannot be described by the $Q=3b^2H\rho$ expression. Following
\cite{Cai2010}, Wei proposed the sign-changeable interaction term
as $Q=q(\alpha \dot{\rho}+3\beta H{\rho})$ for the mutual
interaction \cite{Wei2011,WEI2011}. In this new expression, both
$\alpha$ and $\beta$ are dimensionless constant and
$q=-1-\dot{H}/H^2$ is the deceleration parameter. The key
ingredient is the deceleration parameter $q$ in the interaction
$Q$, and hence the interaction $Q$ can change its sign when our
universe changes from deceleration phase $(q > 0)$ to acceleration
$(q < 0)$. The $\alpha\dot{\rho}$ and $\beta H{\rho}$ terms are
introduced from the dimensional point of view
\cite{Wei2011,chimen1,chimen2}, and one can ignore the
$\alpha\dot{\rho}$ term by considering $\alpha=0$
\cite{Wei2011,chimen1,chimen2}. It is worth noting that the
$Q(\rho,\dot{\rho})$ form has previously been introduced to
describe the mutual interaction between the dark sectors of cosmos
\cite{chimen1,chimen2}. Now, bearing the Firedmann equation in
mind ($H^2\propto\rho$), we find out that the $Q(\rho,\dot{\rho})$
interaction can indeed be written in the $Q(H,\dot{H})$ form. On
the other hand, since $q=q(H,\dot{H})$, the $Q=q(\alpha
\dot{\rho}+3\beta H{\rho})$ interaction can also be written in the
form $Q(H,\dot{H})$, meaning that in the Wei's approach, the
appearance of the deceleration parameter, $q$, in the interaction
term $Q$ is quite acceptable \cite{Wei2011,WEI2011}.

In the present work, we are interested in studying the
cosmological consequences of considering a sign-changeable
interaction term in HDE model.  We shall take $Q=3 b^2 Hq{\rho}$,
where $\rho=\rho_m+\rho_D$ is the total energy density. We will
consider three cutoffs, including the future event horizon, the
Hubble and GO cutoffs, to build the energy density of HDE which
helps us in providing three models for DE. Thereinafter, we
investigate the evolution of the system parameters, such as the
state parameter as well as the deceleration and dimensionless
density parameters, during the cosmos evolution from the matter
dominated era to the current accelerating epoch.

The paper is organized as follows. In section $\textmd{II}$, we
use the future event horizon as the IR cutoff to build the HDE,
and study the evolution of the system's cosmological parameters.
Sections $\textmd{III}$ and $\textmd{IV}$ also include the HDE
models with the Hubble and GO cutoffs, respectively. To have a
better understanding of the system evolution, we plot the
cosmological parameters, such as $q$, the equation of state and
dimensionless density parameters, in terms of the redshift $z$. We
finish with summary and concluding remarks in the last section.
%%%%%%%%%%%%%%%%%%%%%%%%%%%%%%%%%%%%%%%%%%%%%%%%%%%%%%%%%%%%%%%%%%%%%%%%%%%%%%%%%%%
\section{HDE with the sign-changeable interaction with future horizon as IR cutoff}
We consider a homogeneous and isotropic flat FRW universe which is
described by the line element
\begin{eqnarray}\label{frw}
ds^{2}=-dt^{2}+a^{2}\left( t\right) \left[ dr^2
+r^{2}d\Omega^{2}\right],
\end{eqnarray}
where $a(t)$ is scale factor. The first Friedmann equation is
written as
\begin{equation}\label{Fri1}
H^2=\frac{\rho}{3M_p^2},
\end{equation}
where $\rho$ is the total energy density, which satisfy the
conservation law
\begin{eqnarray}\label{cont1}
\dot{\rho}+3H(\rho+p)=0.
\end{eqnarray}
We assume the two dark sectors of the universe, namely
pressureless DM and HDE exchange energy and thus they no longer
satisfy the conservation equations, separately. Instead, they
satisfy the semi-conservation equations as
\begin{eqnarray}\label{conm}
&&\dot{\rho}_m+3H\rho_m=Q,\\
&&\dot{\rho}_D+3H(1+\omega_D)\rho_D=-Q.\label{conD}
\end{eqnarray}
where $\omega_D\equiv{p_D}/{\rho_D}$ is the equation of state
parameter of HDE, and $Q$ denotes the interaction term between DE
and DM. As it is obvious, DE (DM) decays into DM (DE) for $Q>0$
($Q<0$). Following \cite{Wei2011,WEI2011}, we assume the
interaction term has the following form
\begin{eqnarray}\label{Q}
Q = 3b^2Hq(\rho_D+\rho_m),
\end{eqnarray}
where $b^2$ is the coupling constant and $q$ is the deceleration
parameter defined as
\begin{eqnarray}\label{q}
q=-\frac{\ddot{a}a}{{\dot{a}}^2}=-1-\frac{\dot{H}}{H^2}.
\end{eqnarray}
From Eq.~(\ref{Q}), it is obvious that the transition of the
universe from the deceleration $(q>0)$ to an acceleration $(q<0)$
phase, changes the sign of the interaction term $Q$. Defining, as
usual, the dimensionless density parameters as
\begin{equation}\label{Omega}
\Omega_m=\frac{\rho_m}{3M_p^2H^2},\
 \   \   \    \     ~~\Omega_D=\frac{\rho_D}{3M_p^2H^2},
\end{equation}
we can rewrite the first Friedmann equation in the form
\begin{equation}\label{Fri2}
\Omega_m+\Omega_D=1.
\end{equation}
Also, if we take the time derivative of  the ratio of the energy
densities, $r={\rho_m}/{\rho_D},$ we find
\begin{equation}\label{dotr}
\dot{r}={3b^2Hq+6b^2Hqr+3b^2Hqr^2+3Hr\omega_D},
\end{equation}
where we have used Eqs.~(\ref{conm}) and~(\ref{conD}). Solving for
$w_D$, we arrive at
\begin{equation}\label{wD1}
\omega_D=\frac{\dot{r}-3b^2Hq-6b^2Hqr-3b^2Hqr^2}{3Hr}.
\end{equation}
\begin{figure}[htp]
\begin{center}
\includegraphics[width=8cm]{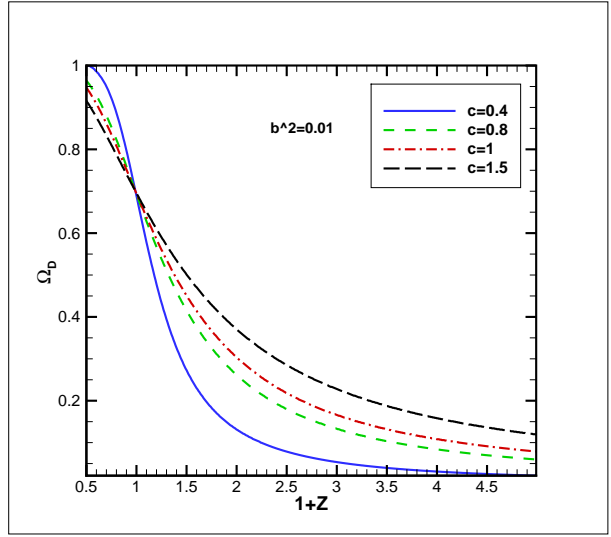}
\caption{Evolution of $\Omega_D$ versus redshift parameter $z$ for
HDE with Future cutoff for a fixed interaction parameter with
coupling interaction $b^2$ but for different values of the constant $c$%Left panel
%corresponds to $\alpha<1$ and the right panel shows the case $\alpha>1$
.}\label{Omega-z1}
\end{center}
\end{figure}
Using Eqs.~(\ref{Omega}),~(\ref{Fri2}), one obtains
\begin{equation}\label{r2}
r=\frac{\Omega_m}{\Omega_D}=\frac{1-\Omega_D}{\Omega_D},
\end{equation}
which implies
\begin{equation}\label{rdot}
\dot{r}=-\frac{\dot{\Omega}_D}{\Omega_D^2}.
\end{equation}
Taking into account the fact that
$\dot{\Omega}_D={\Omega}^{\prime}_{D}H$, after using Eq.
(\ref{rdot}), we can rewrite Eq. (\ref{wD1}) as
\begin{equation}\label {EoS2}
\omega_D=\frac{-{{\Omega}^{\prime}_{D}}/{\Omega_D^2}-3b^2q-6b^2qr-3b^2qr^2}{3r},
\end{equation}
where the dot and the prime stand for the derivative with respect
to the time and $x=\ln{a}$, respectively. Considering the Hubble
radius as the IR cutoff, it is easy to show that $r$ is a constant
\cite{Pavon2005,Shey2011}. Here, we use the future event horizon
as the IR cutoff and so $r$ is no longer a constant \cite{Li2004}.
The future event horizon is defined as \cite{Li2004,Huang2004}
\begin{equation}
R_{h}=a(t)\int_{t}^{\infty}{\frac{dt}{a(t)}}=\frac{c\sqrt{1+r}}{H},
\end{equation}
where $c$ is a constant, and it leads to
\begin{equation}\label{dotfuture}
\dot{R_{h}}={HR_{h}-1}={c\sqrt{1+r}-1}.
\end{equation}
Taking the time derivative of Eq. (\ref{Fri1}), after using
Eqs.~(\ref{conm}),~(\ref{conD}) and~(\ref{Omega}), we obtain
\begin{equation}\label{dotH1}
\frac{\dot{H}}{H^2}=-\frac{3}{2} {\Omega_D}({1+r+\omega_D}).
\end{equation}
Moreover, using~(\ref{dotfuture}) and~(\ref{dotH1}), we obtain
\begin{equation}\label{wD3}
\omega_D=-\frac{1}{3}-\frac
{2\sqrt{\Omega_D}}{3c}-\frac{b^2q}{\Omega_D},
\end{equation}
Combining Eqs.~(\ref{dotfuture}) and~(\ref{dotH1}) we get,
\begin{equation}\label{Omega2}
\frac{{\Omega}^{\prime}_{D}}{\Omega_{D}^{2}}=(1-\Omega_D)\left(\frac{1}
{\Omega_{D}}+\frac{2}{c\sqrt{\Omega_{D}}}-\frac{3b^{2}q}{{\Omega}_D(1-\Omega_D)}\right).
\end{equation}
Substituting Eq. (\ref{dotH1}) into~(\ref{q}), one finds
\begin{equation}\label{q2}
q=\frac{1-\Omega_D-\frac{2\Omega_{D}^{\frac{3}{2}}}{c}}{2+3b^2}.
\end{equation}
Combining this equation with Eq. (\ref{wD3}), we find out that,
for the non-interacting case ($b^2=0$), we have $\omega_D=-1$ in
late time where $\Omega_D\rightarrow 1$ provided we take $c=1$. In
addition, it is easy to check that, for $b^2=0$ and $c=1$, we have
$q=-1$ provided $\Omega_D\rightarrow1$.
\begin{figure}[htp]
\begin{center}
\includegraphics[width=8cm]{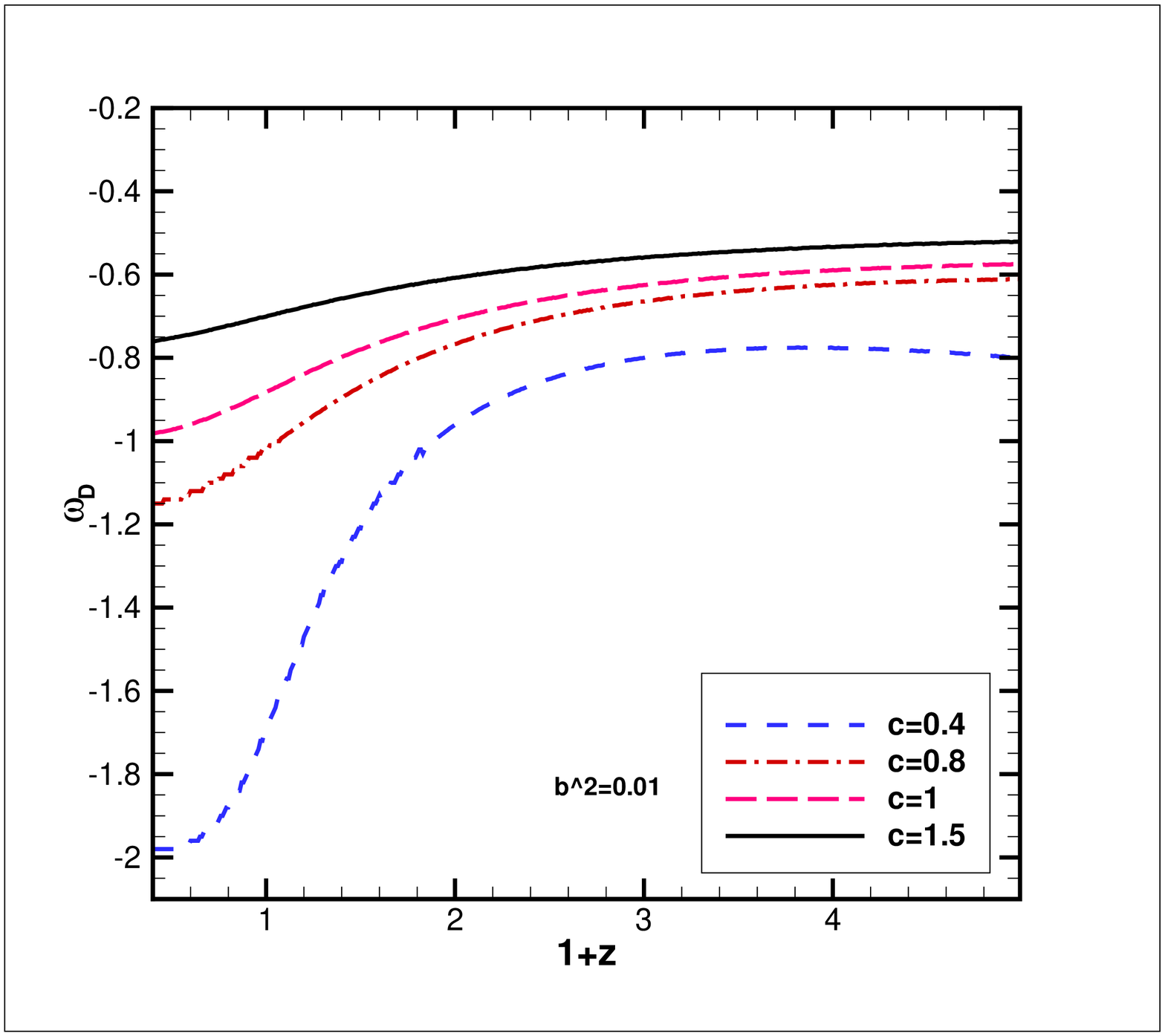}
\includegraphics[width=8cm]{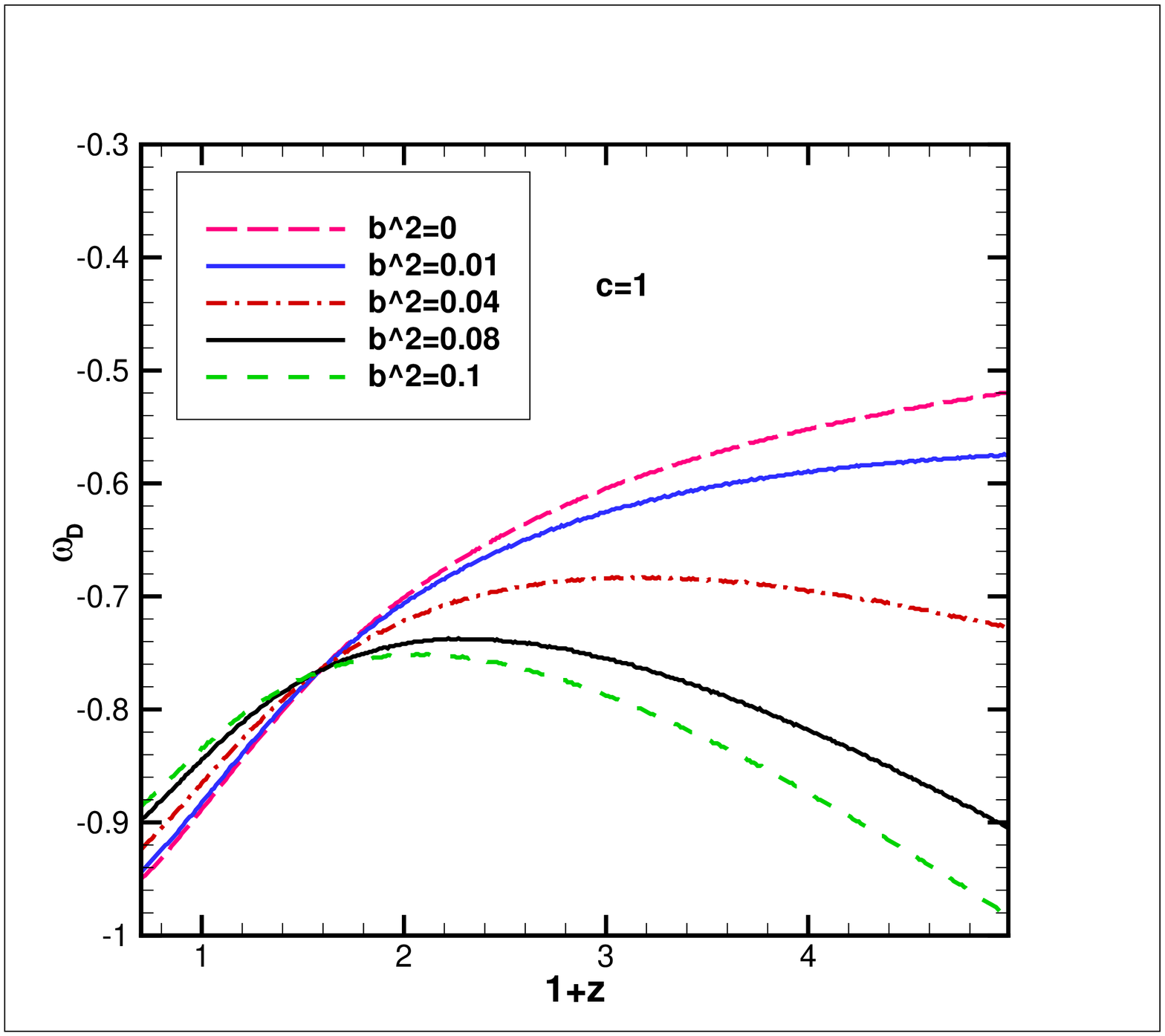}
\caption{The evolution of $\omega_D$ versus redshift parameter $z$
for HDE with Future cutoff. Left panel corresponds to a fixed
couplings between DE and DM and the right panel shows the case in
which $c$ is constant.}\label{EoS-z1}
\end{center}
\end{figure}

The evolution of $\Omega_D$ in terms of redshift parameter ($1+z$)
with respect to the constant $b^2$ is plotted in
Fig.~\ref{Omega-z1}. From this figure we see that, in the early
universe  where $z\rightarrow\infty$ we have $\Omega_D\rightarrow
0$, while at the late time where $1+z\rightarrow 0$ we have
$\Omega_D\rightarrow 1$. It is also apparent that, at fixed $b^2$,
the DE role in the early universe will be highlighted with
increasing the  value of $c$. Additionally, in the long run limit
($1+z\rightarrow 0$), DE plays a more effective role with
decreasing the value of $c$.
\begin{figure}[htp]
\begin{center}
\includegraphics[width=8cm]{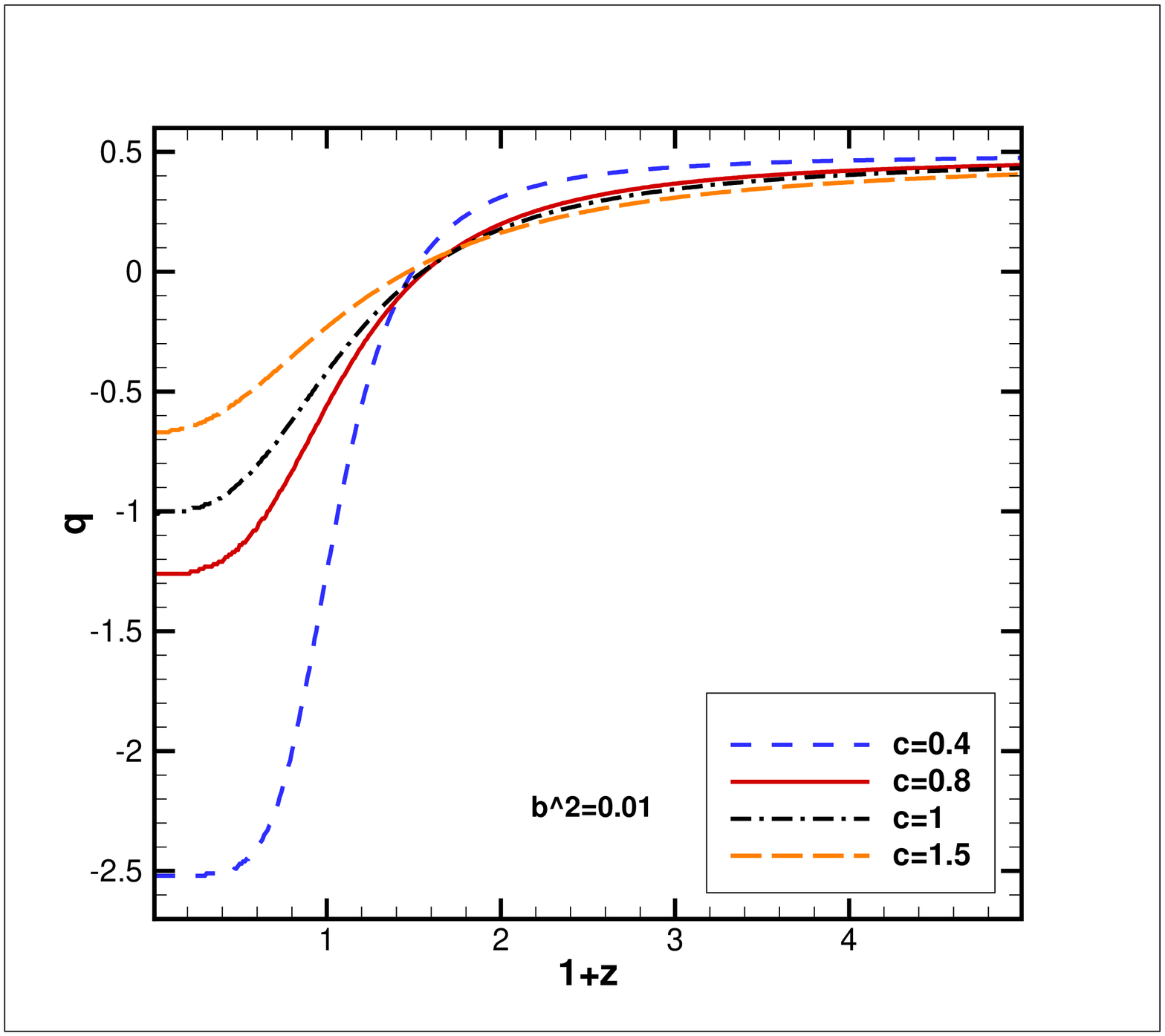}
\includegraphics[width=8cm]{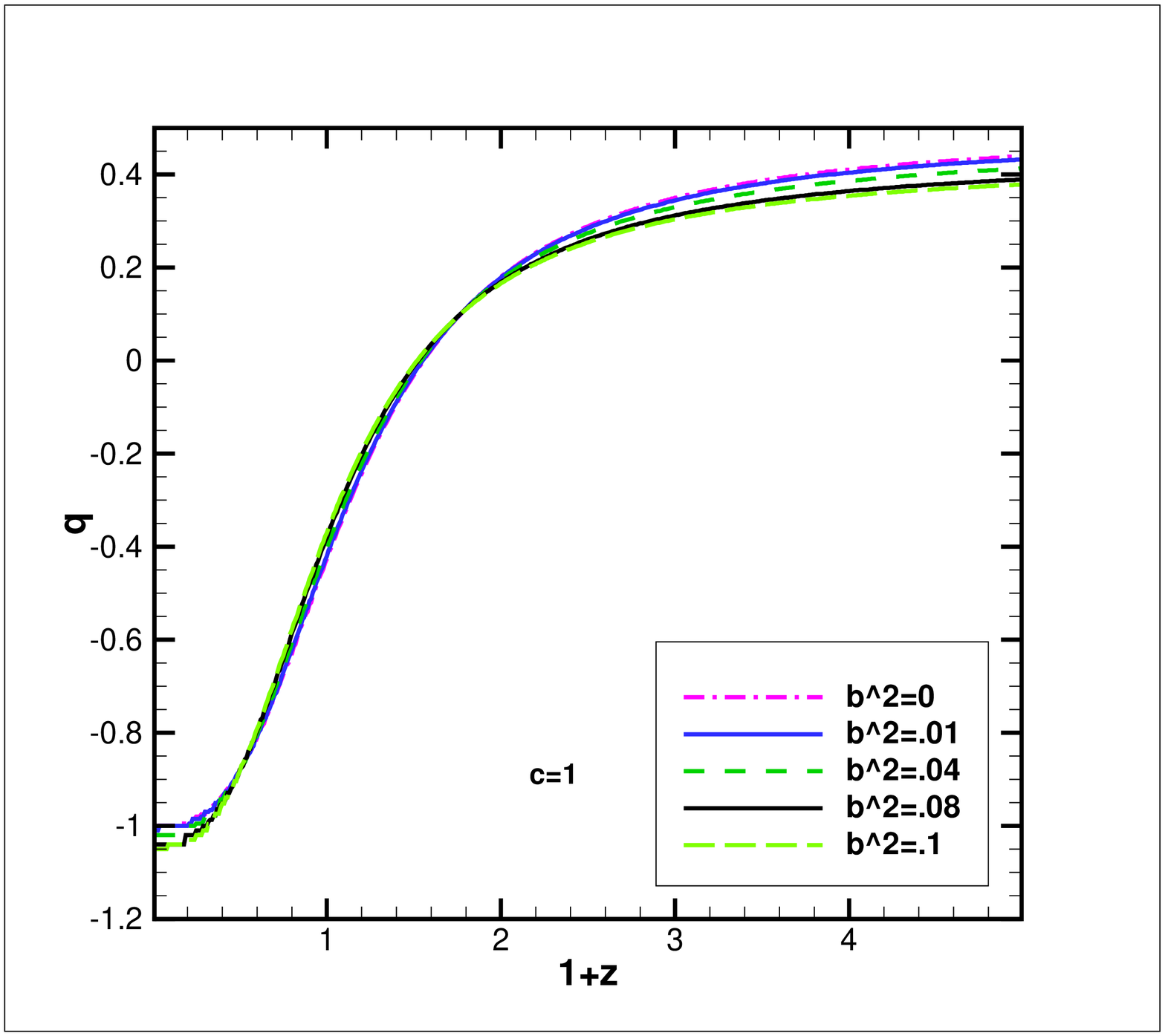}
\caption{The evolution of the  deceleration parameter $q$ against
redshift parameter $z$ for HDE with Future cutoff. Left panel
corresponds to a fixed couplings between DE and DM and the right
panel shows the case $c$ is constant.}\label{q-z1}
\end{center}
\end{figure}
The equation of state (EoS) parameter given in (\ref{wD3}) is also
plotted in Fig.~\ref{EoS-z1} showing that, for a fixed $b^2$, the
EoS parameter can cross the phantom line for $c<1$, and when
$c>1$, we always have $\omega_D>-1$.

The behaviour of the deceleration parameter $q$ is  plotted in
Fig.~\ref {q-z1} showing that there is a deceleration expansion at
the early time followed by an acceleration expansion. In addition,
the transition from the deceleration phase to the acceleration one
happen at $z\approx 0.6$ which is consistent with recent
observations \cite{Daly,Kom1,Kom2}. As Fig.~\ref {q-z1} shows, for
fixed $b^2$, the universe expansion may experience a phantom phase
at current time while $c>1$. Moreover, for two universes with the
same $b^2$ and different values of $c$, the universe with larger
amount of $c$ gets its phase transition point earlier than the
other universe. Finally,  for fixed $c$ and for the current stage
of the universe, this model predicts that the universe expansion
phase may fall in the phantom era as a function of $b^2$.
\begin{figure}[htp]
\begin{center}
\includegraphics[width=8cm]{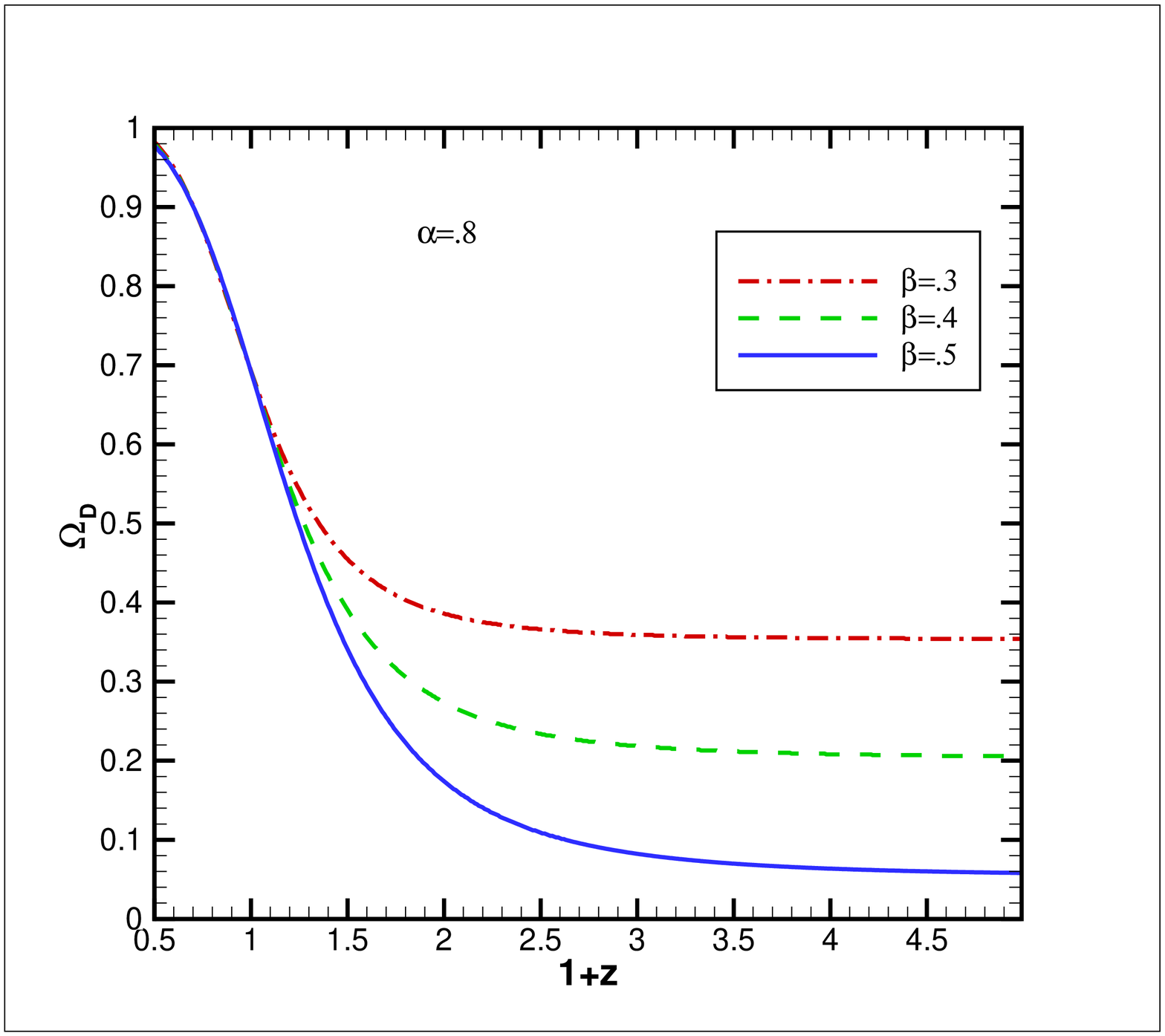}
\includegraphics[width=8cm]{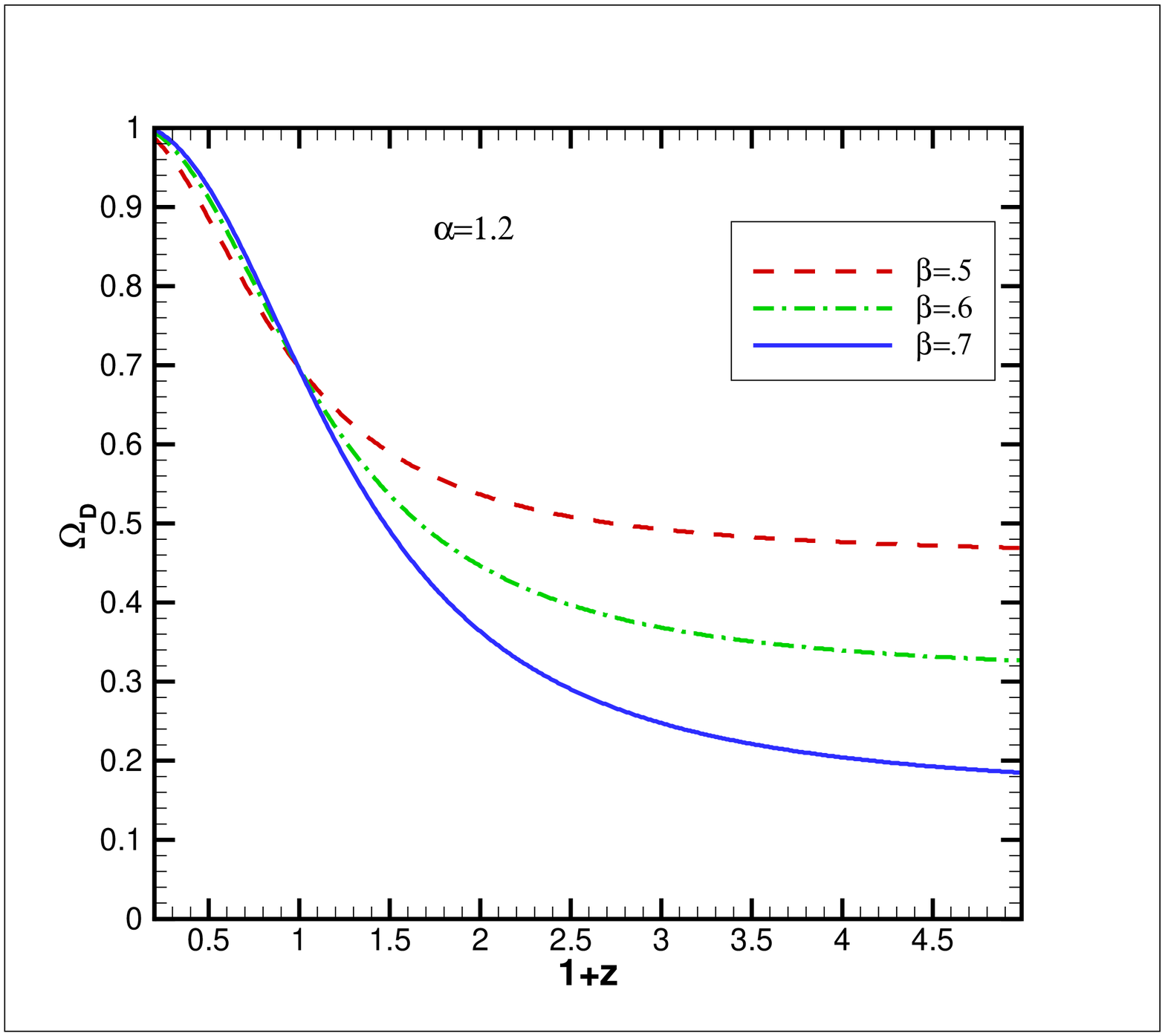}
\caption{The evolution of $\Omega_D$ versus the $1+z$ parameter
for HDE with GO cutoff and $b^2=0\cdot01$. Left panel corresponds
to $\alpha<1$ and the right panel shows the $\alpha>1$
case.}\label{Omega-z2}
\end{center}
\end{figure}
%%%%%%%%%%%%%%%%%%%%%%%%%%%%%%%%%%%%%%%%%%%%%%%%%%%%%%%%%%%%%%%%%%%%%%
\section{HDE with the sign-changeable interaction with Hubble cutoff}
In this section, we use the Hubble radius ($L=H^{-1}$) as IR
cutoff to study the effects of a sign-changeable interaction in
the HDE model. In this case we have $\rho_D=3c^2M_p^2H^2$, where
$c$ is a constant. Finally, bearing the
$\Omega_D={\rho_D}/{(3M_p^2H^2)}$ relation in mind, we find out
$\Omega_D=c^2$ meaning that $c^2$ should meet the $0\leq c^2\leq1$
condition \cite{Ghaffari2015}. It is worth mentioning that, in the
non-interacting model we arrive at $\omega_D=0$ for the HDE with
Hubble radius as the IR cutoff \cite{Li2004}. This implies that in
this case the accelerated expansion of the universe cannot be
achieved \cite{Li2004}. It is also a matter of calculation to
combine the $\rho_D=3c^2M_p^2H^2$ expression with
Eqs.~(\ref{conD}) and~(\ref{dotH1}), and using the $Q = 3 b^2
Hq\rho=3 b^2 Hq(\rho_D+\rho_m)$ relation for the mutual
interaction to get
\begin{equation}\label{EoS4}
\omega_D=-\frac{ b^2 q}{c^2(1-c^2)},
\end{equation}
where
\begin{equation}\label{q3}
q=-1-\frac{\dot{H}}{H^2}=\frac{1-c^2}{2(1-c^2)+3 b^2}.
\end{equation}
It is obvious that in the absence of the interaction term, where $
b^2=0$, we get $w_D=0$ and $q=1/2>0$ which refer to a decelerated
universe. Now, inserting Eq. (\ref{q3}) into Eq. (\ref{EoS4}), we
obtain
\begin{equation}\label{EoS44}
\omega_D=-\frac{ b^2}{c^2[2(1-c^2)+3 b^2]}.
\end{equation}
Clearly, for $2(1-c^2)+3b^2>0$, we have  $\omega_D<0$.

\begin{figure}[htp]
\begin{center}
\includegraphics[width=8cm]{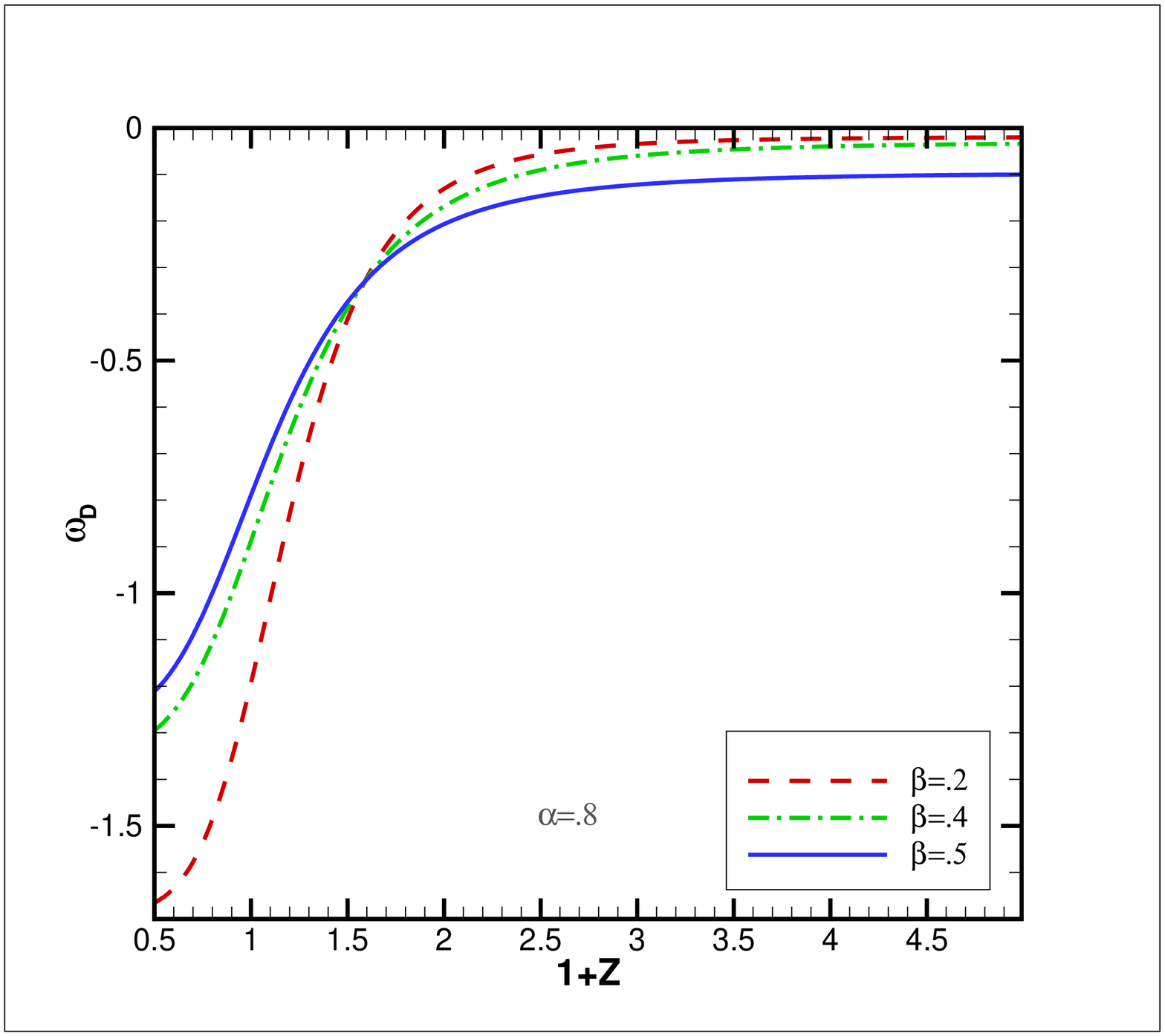}
\includegraphics[width=8cm]{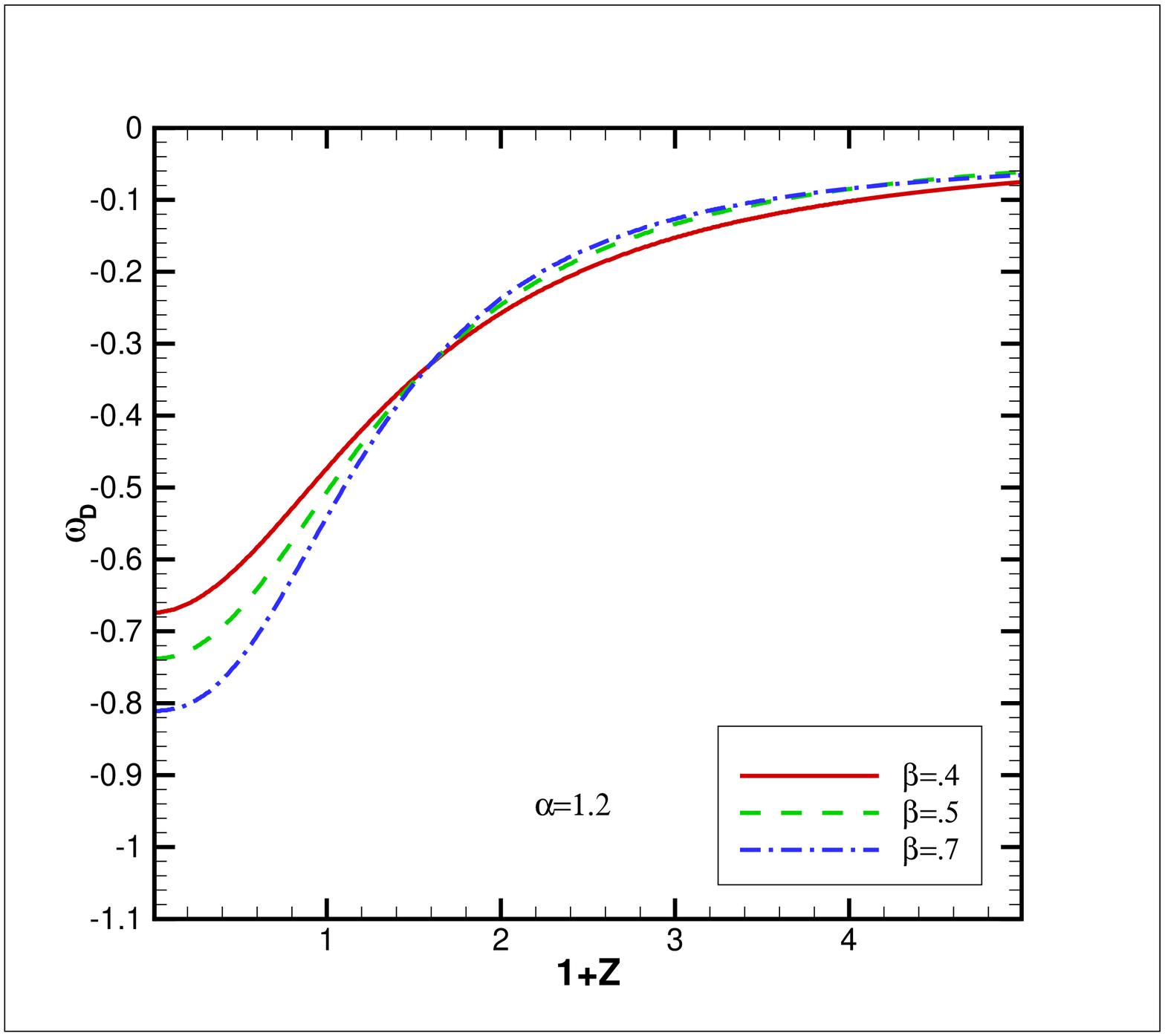}
\caption{The evolution of $\omega_D$ versus the $1+z$ parameter
for HDE with GO cutoff and $b^2=0\cdot01$. Left panel corresponds
to $\alpha<1$ and the right panel shows the $\alpha>1$
case.}\label{EoS-z2}
\end{center}
\end{figure}
%%%%%%%%%%%%%%%%%%%%%%%%%%%%%%%%%%%%%%%%%%%%%%%%%%%%%%%%%%%%%%%%%%%
\section{HDE with the sign-changeable interaction with GO cutoff}
The third cutoff, introduced by Granda and Oliveros (GO)
\cite{Granda2008}, is the generalization of the Ricci scalar
\cite{Gao} in which we have $L=(\beta \dot{H}+\alpha H^2)^{-1/2}$
where $\alpha$ and $\beta$ are constant, and therefore
\begin{equation}\label{GOHDE}
\rho_D=3M_p^2(\alpha H^2+\beta \dot{H}),
\end{equation}
which can also be rewritten
\begin{equation}\label{dotH2}
\frac{\dot{H}}{H^2}=\frac{\Omega_D-\alpha}{\beta}.
\end{equation}
Taking the time derivative of Eq. (\ref{Fri1}) and using
Eqs.~(\ref{conm}),~(\ref{conD}),~(\ref{Omega}) and~(\ref{Fri2}),
we reach
\begin{equation}\label{HDEdot}
\frac{\dot{\rho}_D}{3M_p^2H^3}=\frac{2\dot{H}}{H^2}+3(1-\Omega_D)-3b^2q.
\end{equation}
From Eq.~(\ref{Omega}), we obtain
\begin{equation}\label{Omega3}
\dot{\Omega}_D=\frac{\dot{\rho}_D}{3M_p^2H^2}-2\Omega_D\frac{\dot{H}}{H}.
\end{equation}
Combining Eqs.~(\ref{dotH2}) and~(\ref{HDEdot}) with
(\ref{Omega3}) to get
\begin{equation}\label{Omega4}
{\Omega}^\prime_D=(1-\Omega_D)
\left[3+\frac{2}{\beta}(\Omega_D-\alpha)\right]-3b^2q.
\end{equation}
Inserting Eqs.~(\ref{GOHDE}),~(\ref{dotH2}) and~(\ref{HDEdot})
into (\ref{conD}), and combining Eqs.~(\ref{dotH2}) and~(\ref{q}),
we find the EoS  and deceleration parameter as
\begin{equation}\label{EoS5}
\omega_D=-\frac{1}{\Omega_D}\left[1+\frac{2}{3}\left(\frac{\Omega_D-\alpha}{\beta}\right)\right],
\end{equation}
\begin{equation}\label{q4}
q=-1-\frac{\dot{H}}{H^2}=-1-\frac{\Omega_D-\alpha}{\beta}.
\end{equation}
The evolution of HDE density parameter ($\Omega_D$) in terms of
redshift parameter ($1+z$), whenever $b^2=0\cdot01$, is shown in
Fig.~\ref{Omega-z2}. As it is obvious from Fig.~\ref{Omega-z2}, we
have $\Omega_D\rightarrow 0$ and $\Omega_D\rightarrow 1$ in the
early universe ($1+z\rightarrow\infty$) and the $1+z\rightarrow 0$
limit, respectively.

Since Eq.~(\ref{EoS5}) indicates that we have $\omega_D=-1$ while
$\Omega_D\rightarrow1$ and $\alpha=1$ simultaneously, we plot the
$\alpha<1$ and $\alpha>1$ cases separately in Fig.~\ref{EoS-z2}.
As it is obvious from this figure, the model can show a
phantom-like behavior ($\omega_D<-1$) for $\alpha<1$.
Additionally, for the $\alpha>1$ case, $\omega_D$ is always larger
than $-1$.
\begin{figure}[htp]
\begin{center}
\includegraphics[width=8cm]{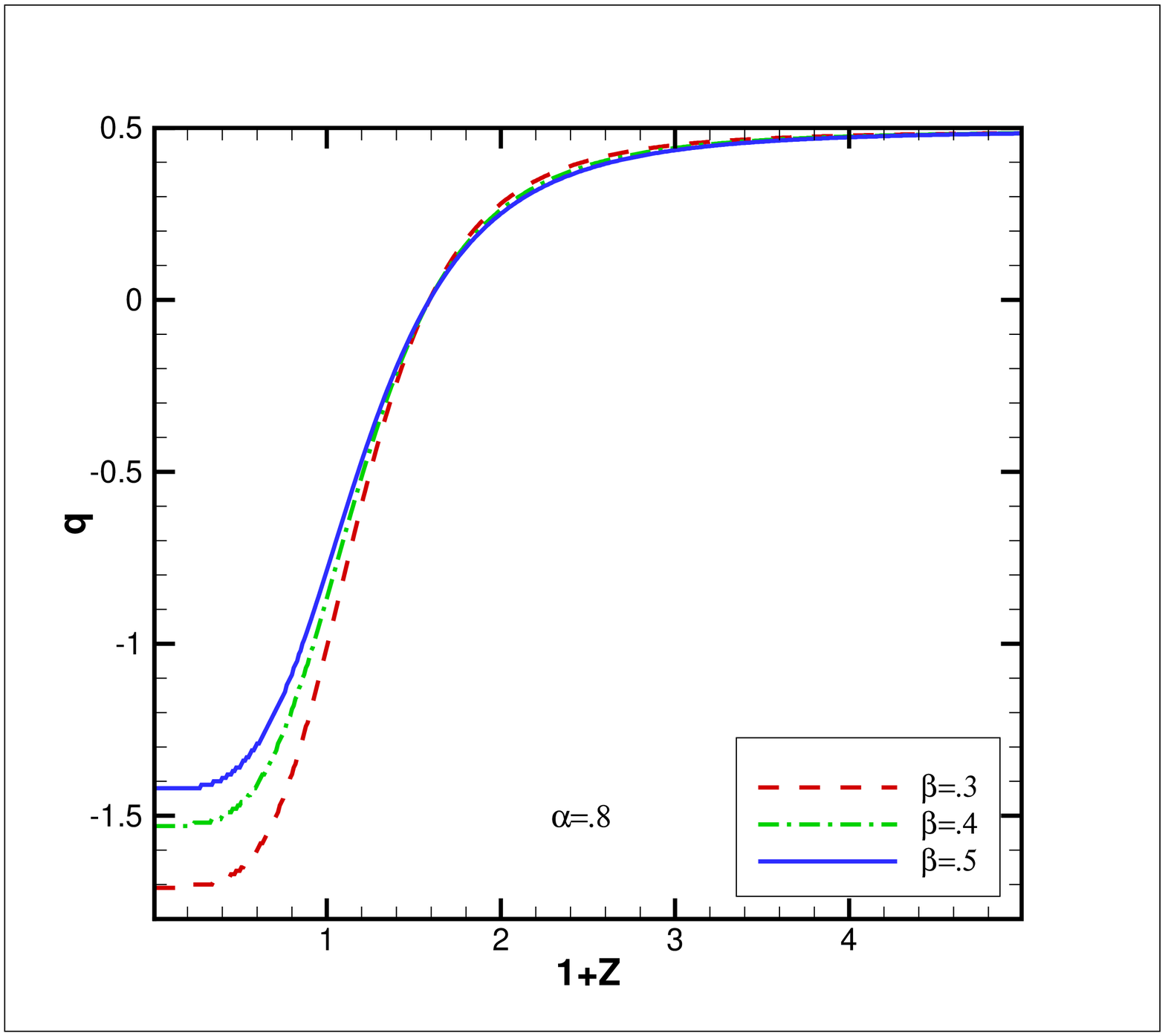}
\includegraphics[width=8cm]{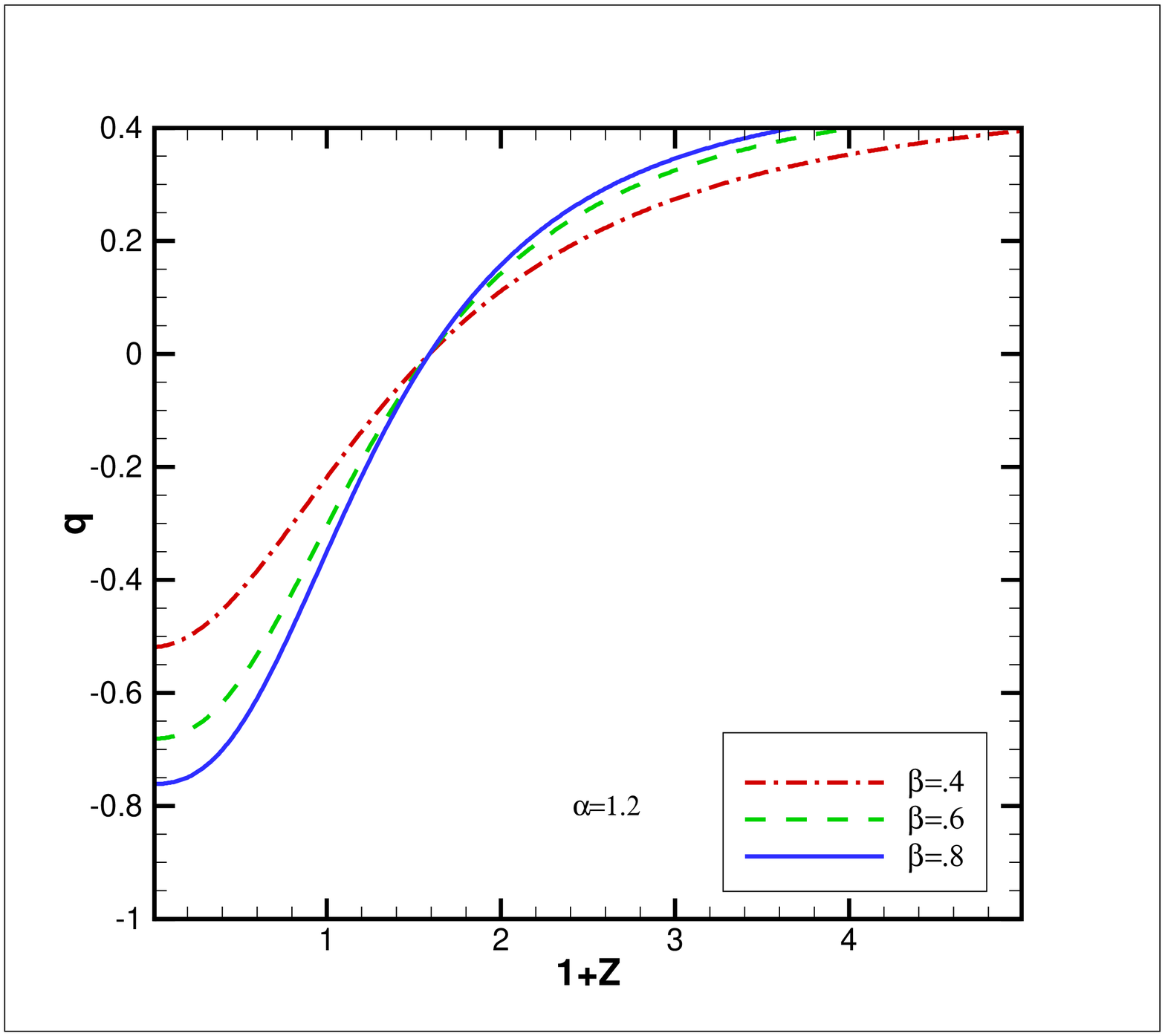}
\caption{The evolution of $q$ versus the $1+z$ parameter for HDE
with GO cutoff and $b^2=0\cdot01$. Left panel corresponds to
$\alpha<1$ and the right panel shows the case
$\alpha>1$.}\label{q-z2}
\end{center}
\end{figure}
The behavior of $q$ is also shown in Fig.~\ref{q-z2}. It is
obvious that the universe experiences a transition from a
deceleration phase to an acceleration one around $z\approx
0\cdot6$, compatible with observation \cite{Daly}. It is also
useful to note that Eq.~(\ref{q4}) shows $q\rightarrow-1$ while
$\Omega_D\rightarrow1$ and $\alpha=1$.
\section{Summary and concluding remarks}
Throughout this paper, we considered a universe filled by dark
matter and dark energy, in which there is a sign-changeable
interaction [$Q = 3b^2Hq(\rho_D+\rho_m)$] between the dark
sectors. Considering the future event horizon, the Hubble radius
and GO as the system IR cutoffs, we followed the HDE idea to
obtain three models for dark energy.

At first, we used the future event horizon as the IR cutoff to
construct the HDE density profile. In addition, we studied the
evolution of density parameter, the deceleration parameter and the
EOS parameter. Our study shows that the model is compatible with
some recent observations, and it predicts a transition from the
deceleration phase to an acceleration one at around
$z\approx0\cdot6$. Our check indicates that, for fixed $b^2$, we
have $\omega_D>-1$ for $c>1$. We also saw that, for fixed $b^2$,
the phantom-like behavior ($\omega_D<-1$) may be obtained for
$c<1$.

Moreover, we focused on the Hubble and the GO cutoffs and found
out that these cutoffs may also lead to HDE models compatible with
observations. In fact, we saw that, unlike the non-interacting HDE
model with the Hubble cutoff \cite{Li2004}, it is possible to have
$\omega_D<0$ and $q<0$ in the interacting case. In addition, our
study shows that the model with the GO cutoff can predict the
universe transition from a deceleration phase to an acceleration
one at around $z\approx0\cdot6$. We finally found out that the EOS
of the model with GO cutoff may also cross the phantom line.
%%%%%%%%%%%%%%%%%%%%%%%%%%%%%%%%%%%%%%%%%%%%%%%%%%%%%%%%%%%%%%%%%%%%%%%
\acknowledgments{We also thank Shiraz University Research Council.
This work has been supported financially by Research Institute for
Astronomy \& Astrophysics of Maragha (RIAAM), Iran.}
%%%%%%%%%%%%%%%%%%%%%%%%%%%%%%%%%%%%%%%%%%%%%%%%%%%%%%%%%%%%%%%%%%%%%%%

\end{document}